\documentclass{ws-ijmpcs}

\begin{document}

\markboth{F. Margaroli}
{TOP PHYSICS AT THE TEVATRON AND LHC}

%
\catchline{}{}{}{}{}
%

\title{MEASUREMENT OF TOP QUARK PROPERTIES AT THE TEVATRON AND LHC}

\author{Fabrizio Margaroli\footnote{On behalf of the CDF, D0, CMS, ATLAS collaborations.}}

\address{Physics Department, Sapienza University of Rome \& INFN, \\
Piazzale Aldo Moro 5 Rome, 00185, Italy\\
fabrizio.margaroli@roma1.infn.it}

\maketitle


\begin{abstract}
Almost two decades after its discovery at Fermilab's Tevatron collider experiments, the top quark is still under the spotlight due to its connections to some of the most interesting puzzles in the Standard Model. The Tevatron has been shut down two years ago, yet some interesting results are coming out of the CDF and D0 collaborations. The LHC collider at CERN produced two orders of magnitude more top quarks than Tevatron's, thus giving birth to a new era for top quark physics.  While the LHC is also down at the time of this writing, many top quark physics results are being extracted out of the 7\,TeV and 8\,TeV proton proton collisions by the ATLAS and CMS collaborations, and many more are expected to appear before the LHC will be turned on again sometime in 2015. These proceedings cover a  selection of recent results produced by the Tevatron and LHC experiments.
\end{abstract}

\ccode{PACS numbers:11.15.Ex, 12.15.-y, 14.65.Ha, 14.80.Ec, 14.80.Fd}

\section{Introduction}	

The discovery of the top quark\,\cite{Abe:1995hr,Abachi:1995iq} during the 1.8\,TeV center-of-mass energy collisions of protons and antiprotons at the Tevatron was probably the single major achievement to stem from the effort of the CDF and D0 collaborations. While the existence of top quarks was expected, the community was surprised by the very large value that has been measured for its mass, approximately 175\,GeV; the top quark stands as the most massive elementary particle even after the discovery of the Higgs boson\,\cite{Chatrchyan:2012ufa,Aad:2012tfa}. This large value prompted a number of questions: why its mass sits at the electroweak scale, being some 40 times larger than the heaviest fermion mass? Why is the Yukawa coupling of the top quark so compatible (by 0.5\%) with one and does that mean the top quark plays a special role in the electroweak symmetry breaking (ESB) mechanism? Given the recently known Higgs mass, more puzzles arise: the top quark large mass causes the so-called hierarchy problem, and top quarks participate in two of its most popular solution - SUSY and composite Higgs - where heavy bosonic or fermionic partners of the top quark would appear at energy scales that should be accessible by the LHC. While the large top quark mass poses numerous theoretical challenges to the Standard Model (SM), it also provide the unique opportunity to study a bare quark as the top quark lifetime ($\tau \sim 10^{-25}$\,s) is much shorter than the hadronization time ($1/\Lambda_{QCD} \sim 10^{-24}$\,s). The top quark is thus the only quark whose properties can be probed directly.

The Tevatron run at 1.96\,TeV in its eight years of operations increased the available top quark dataset by two orders of magnitude with respect to the sample needed for discovery, thanks to the increased cross section at higher center-of-mass energy, approximately 7\,pb\,\cite{Baernreuther:2012ws} and especially the much larger integrated luminosity $L \simeq 10$\,fb$^{-1}$. In its two years of high energy operations, the proton-proton collisions at the LHC allowed a further increase in the available data on top quarks by two more orders of magnitude, thanks mostly to the much larger cross section, $\sim 160 (230)\,$pb at 7(8)\,TeV collisions. The integrated luminosity at the LHC corresponds to 5\,fb$^{-1}$ of data during the 7\,TeV run and $\simeq 19.6$\,fb$^{-1}$ at 8\,TeV collisions. 

There are notable differences among top quark production at the Tevatron and at the LHC. At the Tevatron, in order to produce top quarks, the incoming partons participating in the collision have to carry a very large fraction $x$ of the proton momentum, where the up and down quark parton density functions (PDF) largely dominate over the others. For this reason, at the Tevatron top quarks are mostly produced through quark-antiquark annihilation. On the other hand, due to the higher LHC beam energies - that translate into a lower $x$ needed to produce top quarks - and the fact that it is harder to extract an antiquark from a proton, top quarks are mostly produced through gluon-dominated initial states at the LHC. Pair production induced by quantocromodynamics (QCD) is more abundant than single top quark production induced by electroweak (EWK) interactions and due to the more striking signature, event selections typically leave clean samples, most properties are thus measured in events where top quarks are pair-produced. Details about top quark production have been amply discussed by the previous speaker so we leave its discussion to the corresponding proceedings\,\cite{Reinhard}. It should be noted that the ``historical" distinction of QCD top quark pair production and EWK single top quark production carries a decreasing weight in an era where experimenters are interesting in mixed QCD-EWK processes such as $t \bar t W/Z$ or $t \bar t H$, and single top quark production is of interest both when it appears together with other quarks, or photons, or heavy bosons. 

\section{The crucial role of the top quark mass}

\begin{figure*}
  \begin{center}
    \includegraphics[width=10cm]{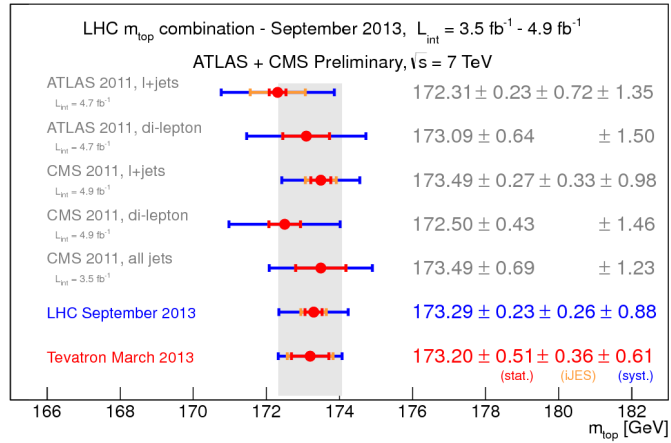}
    \caption{\label{fig:combo} Summary of the most precise top quark mass measurement from the LHC and Tevatron experiments. The central value as measured by the experiments at both colliders is impressively similar, considering that only a fraction of systematic uncertainties are correlated between the measurements. }
  \end{center}
\end{figure*}

The precise knowledge of the top quark mass is of the utmost importance for the understanding of the SM and the exploration of what lies beyond.  In particular, global fits of this and other measurements of electroweak SM parameters provide a stringent prediction for the most likely region for the Higgs mass; the Higgs boson has been discovered at the edge of the 68\% confidence level (CL) of these predictions\,\cite{Baak:2012kk}. Now that the Higgs mass is known, its value is fed into the same fit to assess SM self-consistency and constrain beyond SM scenarios (BSM). The knowledge of top quark and Higgs boson masses has recently been used to assess the vacuum stability, with important cosmological consequences\,\cite{Degrassi:2012ry}. The broad impact the top quark mass has in physics commands the need for ever-increasing precision.
The ATLAS collaboration released its most sensitive result in a new analysis that measures simultaneously the top quark mass and both light and $b$-jet energy scale\,\cite{ATLAS-CONF-2013-046}, the latter two being the two largest systematic uncertainties that affect the mass determination. This measurement is now limited by other systematics, most notably the ones involved in the physics modeling of the signal itself. The top quark mass measurements performed by the Tevatron, CMS and ATLAS are all consistent with the value of 173\,GeV within a few hundreds MeV\,\cite{TevCombo,ATLAS-CONF-2013-046,Chatrchyan:2013xza} and are summarized in Fig.\,\ref{fig:combo}. The most precise results make use of the full reconstruction of the top quarks by recombinations of their daughter particles. This technique allows also to probe the hypothesis that the top quark width $\Gamma_t$ differs from SM predictions, possibly because of exotic top quark decays. The CDF collaboration performed such measurements, setting constraints on generic exotic decays\,\cite{Aaltonen:2013kna}. The main downsides of the mass measurements that make use of full top quark invariant mass reconstruction are the large dependence on the precise knowledge of the jet energy scale, and the intrinsic uncertainty on the definition of the mass of a colored particle. The former point has been addressed by a recent CMS measurement that makes use of the lifetime of the $B$-hadron stemming from the hadronization of the $b$-quark decaying from the top quark\,\cite{CMS-PAS-TOP-12-030}. In fact, the mean of the $B$ meson boost depends linearly on the mass of the top quark. The measurement's sensitivity to the jet energy scale uncertainty is greatly reduced, although at the cost of increasing the sensitivity to other systematic uncertainties. The second issue is addressed by measuring the properly defined top quark ${\it pole}$ mass  intersecting the measurement of the cross section for $t \bar t$ production with its theoretical prediction, as a function of the top quark mass. CMS performs the most precise such measurement finding very good consistency with the results from direct top quark mass measurements\,\cite{CMSpole}. It will be extremely interesting to see such measurements performed using the full 2012 LHC dataset and the most up-to-date theoretical calculations for $t \bar t$ cross section.

\section{Top quark and tests of conservations laws}

The measurement of the difference between the top quark and the antitop quark mass is an interesting analysis that stems naturally out of the measurement of the top quark mass in top quark pair events. It probes directly the CPT theorem for the first time for ``naked" quarks. Semileptonic $t \bar t$ system decays are used, so to tag the top or antitop quark according to the charge of the daughter lepton, and reconstruct its mass accordingly. The CMS collaboration produced the single most precise such result as of today\,\cite{CMS PAS-12-031}. In the measurement of the differences between $m_{t}$ and $m_{\bar t}$ most systematics naturally cancel out, thus enabling vewry high precision: $m_t - m_{\bar t} = -0.27 \pm 0.20 (stat.) \pm 0.12 (syst.)$\,GeV; the relative uncertainty corresponds to $\sigma_{\frac{m_t - m_{\bar t}}{m_t}} \sim 1.3 \times 10^{-3}$. Recently, the CMS collaboration also produced the first test of baryon number (BN) conservation in the top quark system\,\cite{CMS-PAS-B2G-12-023}. In particular, a four-fermion effective lagrangian is assumed that would give rise to decays such as $t \to b c \mu$ and $t \to b u e$. Top quark pair events are used for this purpose, where one top quark decays according to the SM, and the other violates BN conservation. The requirement of a charged lepton without the accompanying neutrino greatly suppresses SM backgrounds, thus allowing to set stringent limits on BN violation: $BR(t \to b c \mu) < 16 \times 10^{-4}$.

\section{The last man standing: asymmetry in $\mathrm{t \bar t}$ quark production}

\begin{figure*}
  \begin{center}
    \includegraphics[width=6.5cm]{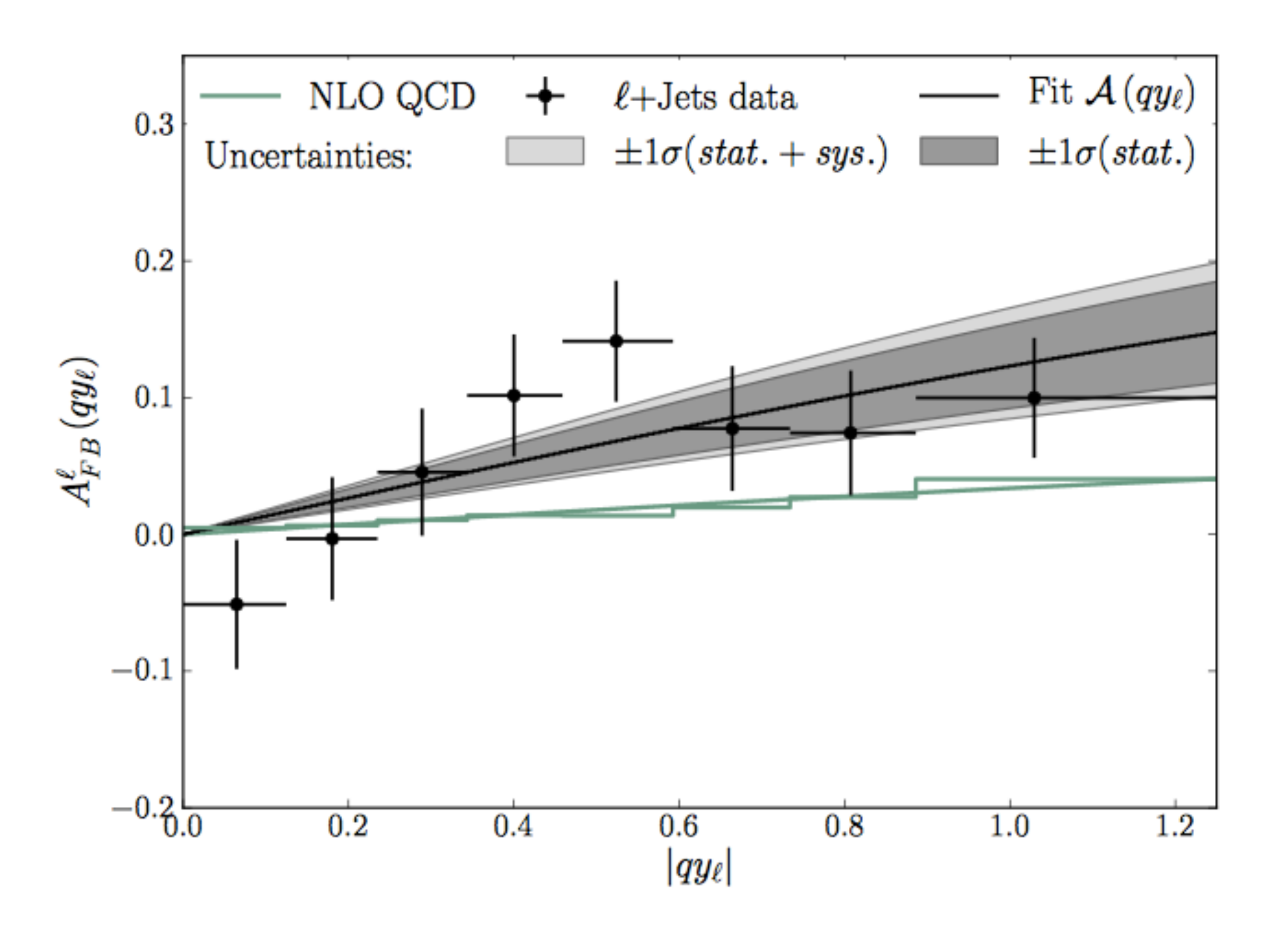}
    \includegraphics[width=6cm]{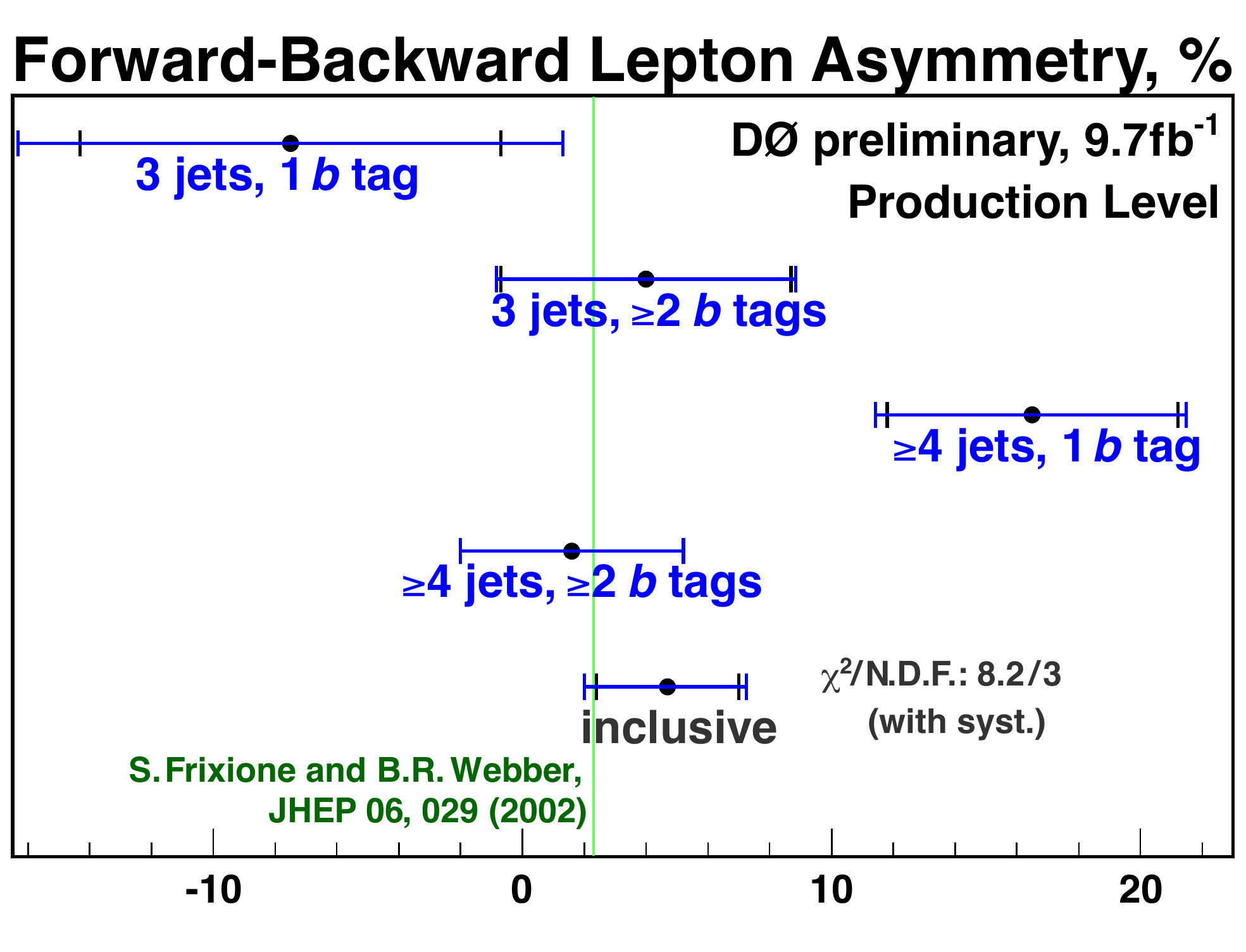}
    \includegraphics[width=6.4cm]{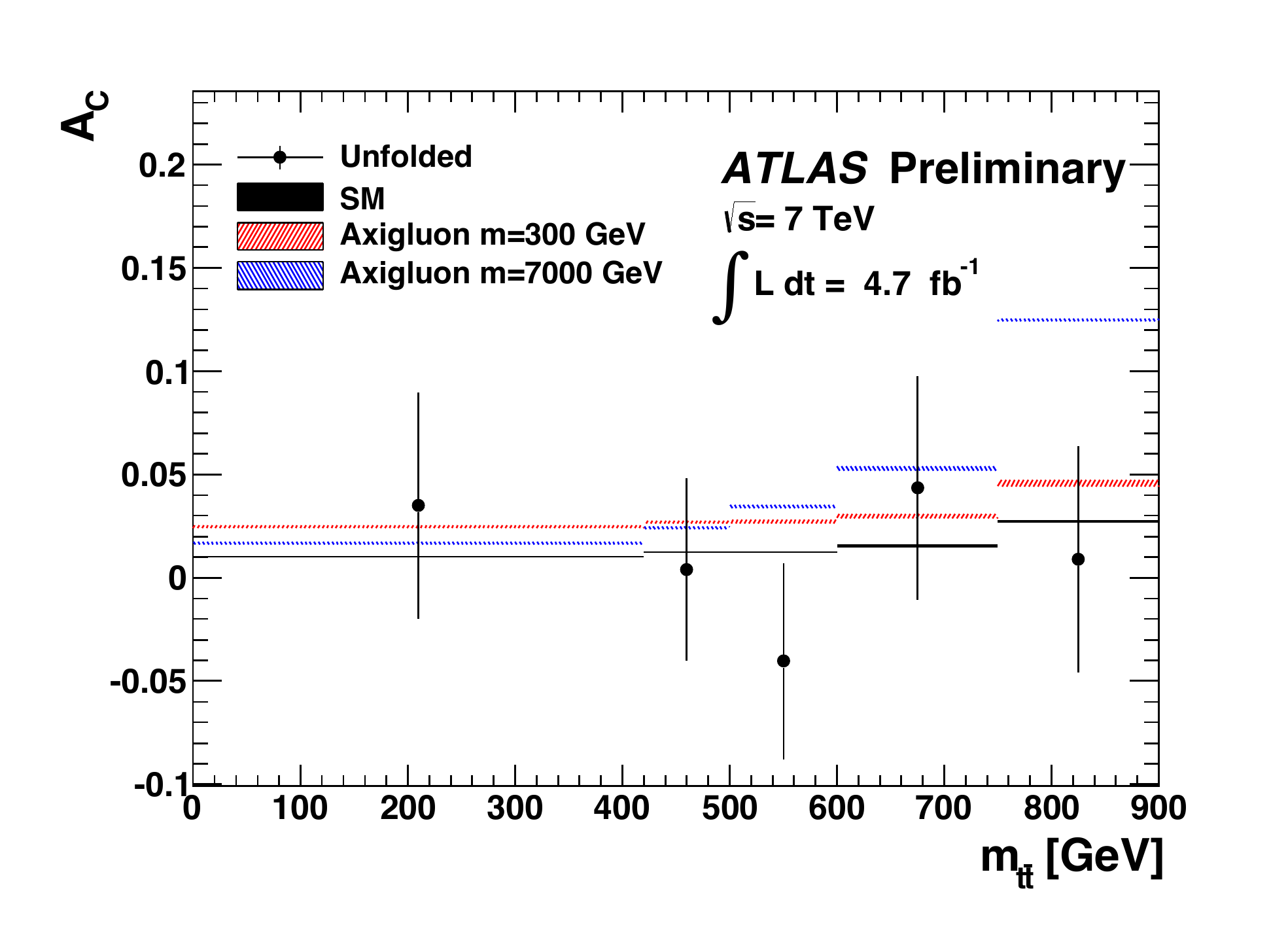}
    \includegraphics[width=6cm]{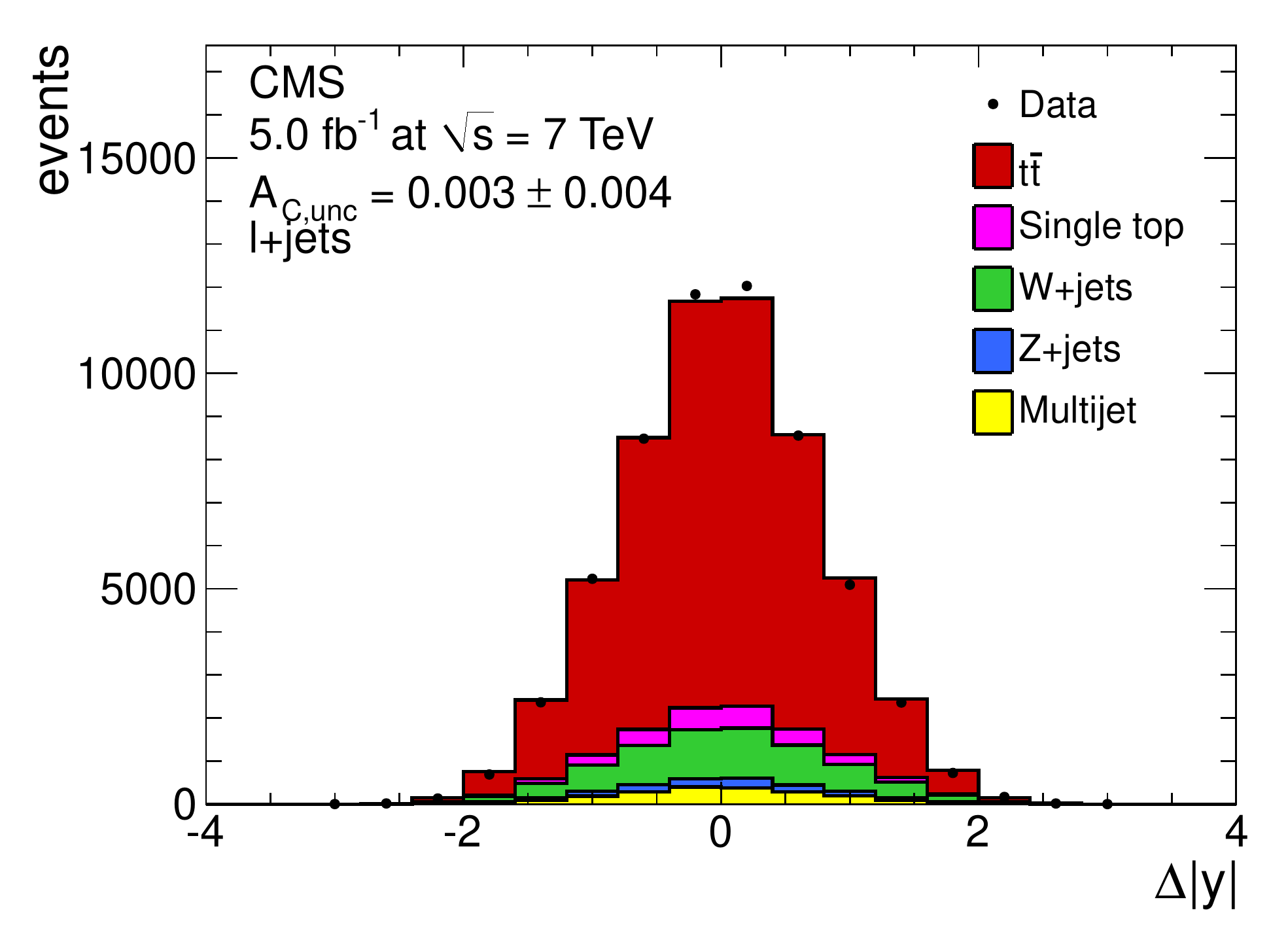}
    \caption{\label{fig:afbs} Distributions of the the lepton (from top quarks) asymmetry in $t \bar t$ decays on the top row, and of the charge asymmetry in the bottom row. The collaborations studies also the dependence fo these observables as a function of other event properties.}
  \end{center}
\end{figure*}

At next-to-leading-order (NLO), QCD predicts the top quark to be emitted preferentially in the direction of flight of the proton beam, while the top antiquark in the direction of the antiproton beam. This charge asymmetry comes mainly from the interference between $q \bar q \to t \bar t$ tree diagram with the NLO box diagram, and from the interference of initial and final state radiations $q \bar q \to t \bar t g$.
The CDF and D0 experiments observed over the years a forward-backward asymmetry (AFB) in top-antitop events that exceeded the SM expectations of a small but non-zero effect. With more data accumulating, CDF observed in 2011 for the first time a $3\sigma$ effect in the dependence of AFB as a function of the $t \bar t$ invariant mass\,\cite{Aaltonen:2011kc}. A large number of new physics models that would explain such a large effect, while predicting new particles that should have appeared at a scale accessible at the Tevatron and/or at the LHC, have been proposed. The experimentalists proceeded simultaneously in the direction of performing new AFB measurements at the Tevatron, and the related-but-not-identical charge asymmetry measurement at the LHC, while searching for additional new physics manifestations. Theoreticians investigated possible effects whose absence in the current calculations would have lead to an underestimated asymmetry (recalculation of NLO EWK effects, and of perturbative and non-perturbative QCD effects).
The latest results produced by the CDF and D0 experiment return a less significant deviation from SM predictions as the new theoretical computation predict larger values and more data and more final states were used\,\cite{Aaltonen:2013vaf,D0 CONF-6394,Abazov:2013wxa}. Both collaborations investigated various observables and their correlations to establish agreement with the SM. Now that the Tevatron has extracted all possible information related to the top quark forward-backward asymmetry, the next word in the chapter rests in the LHC data. As of today, ATLAS and CMS performed differential measurements of the $t \bar t$ asymmetry as a function the invariant mass, the transverse momentum and the rapidity of the $t \bar t$ system\,\cite{CMS-PAS-TOP-12-004,Chatrchyan:2012cxa,ATLAS-CONF-2013-078,ATLAS-CONF-2012-057}. They both observe consistency between the measurements and the predictions in a broad range of observables and kinematic regimes. Some of these observables and their comparison with theoretical predictions are shown in Fig.\,\ref{fig:afbs}. Still, more data is needed - and already available - to draw definitive statements.

\section{Top and charged Higgs}

A rather minimal extension of the SM, with a limited number of new parameters to adjust to experimental data, consists in the two-Higgs doublets model (2HDM). Different 2HDM alternatives have been suggested, with MSSM the most notable example. The most striking prediction of this set of models is the existence of charged Higgs bosons. In particular, charged Higgs bosons lighter than the top quark, could appear in the top quark decay chain $t \to H^{\pm} b$. The branching ratio of these charged Higgses would be largest when decaying into the heaviest fermions, i.e. a $c \bar s$ pair or a $\tau \nu_{\tau}$ pair (charged conjugates processes are implied) where the choice of the former or the latter would be determined by internal parameters of the model. 
The most recent LHC results use the 7\,TeV dataset for these searches. The ATLAS collaboration performed a search of events where the charged Higgs boson would decay favorably to a quark pair\,\cite{Aad:2013hla}. The invariant mass of the charged Higgs boson can thus be completely reconstructed, although at the cost of accepting combinatorial background. A scan of the diet mass compatible with the charged Higgs hypothesis gives back the null results, which can be translated into upper limits on $BR(H^+ \to c \bar s)  \sim 2\%$ depending on the assumed charged Higgs mass. Both the ATLAS and CMS collaborations explored the scenario where the Higgs boson decays most favorably to heavy leptons\,\cite{Aad:2013hla,CMS-PAS-HIG-12-052}. Both leptonic and hadronic decays of the tau lepton are considered; the collaborations use a number of observables sensitive to such exotic decays. In absence of an excess of events consistent with the signal hypothesis, upper limits are set at 95\% CL on the charged Higgs branching ratio, that can be translated into limits on 2HDM parameter space. These limits are of the order of $\sim 2\%$ level, about one of order of magnitude stricter than previous results from the Tevatron collider experiments. A charged Higgs heavier than a top quark, could be produced directly and subsequently decay into a top-antibottom pair. This final state resembles closely single top $s$-channel production, and is very difficult at the LHC due to the very large $W$+heavy flavor jets background. The most recent result on the subject has been produced by the D0 collaboration several years ago\,\cite{Abazov:2008rn}.

\section{Top quark and flavor}

The Cabibbo Kobayashi Maskawa (CKM) matrix elements can be inferred indirectly form the knowledge of the other elements, by exploiting the unitarity of the matrix and its $3 \times 3$ structure. New physics could violate these assumptions in several ways, the most obvious being the possible existence of a fourth generation of quarks. Electroweak single top quark production cross section $\sigma_t$ is proportional to $V_{tb}^2$, thus allowing indirect extraction of this CKM parameter through comparison with the predicted $\sigma_t$. The large number of $\sigma_{t+X}$ measurements produced at the Tevaton and LHC ($X=q,b,W$) have been directly translated into measurements of $V_{tb}$. All results are consistent with the SM predictions of $V_{tb} \simeq 1$.

Processes that are induced by flavor-changing neutral currents (FCNC) in top quark production or decay are extremely small according to SM computation, with decay rates of the order of $10^{-10}$ or below. New physics scenarios such as R-parity violating SUSY, top color technicolor, etc. could enhance FCNC rates by several orders of magnitude, thus making them accessible using current experimental data. Setting stringent limits on the rare decays of a particle - the top quark - that is already rarely produced, requires the wealth of LHC top quark data. FCNC couplings such as $ugt$ or $cgt$ can be probed through the production of a single top quark and no additional particles, through the process $qg \to t \to Wb \to \ell \nu b$ where $(q=u,c)$. Decay products are relatively boosted with respect to the dominant $W+b$ background, and the kinematics of the event is fully reconstructed; several such observables are fed into a Boosted Decision Tree (BDT) to enhance sensitivity, and the limits on such FCNC rates are set via a likelihood scan of the the aforementioned multivariate discriminant. The analysis is more sensitive to $ugt$ rather than $cgt$ coupling due to the up quark PDF being larger than the charm quark PDF in the proton. The ATLAS collaboration sets 95\% confidence level upper limits are $BR(t \to ug)<3.1 \times 10^{-5}$ and $BR(t \to cg)<1.6 \times 10^{-4}$ using 14.2\,fb$^{-1}$ of data\,\cite{ATLAS-CONF-2013-063}. Experimental upper limits on couplings such as $tZq$ can be set analyzing top quark pair production where one top quark decays to a $Z$ boson and a light quark, in events such as $t \bar t \to b W q Z \to multileptons$. CMS performed such an analysis on the full 2012 dataset corresponding to 19.5\,fb$^{-1}$ where the kinematics can be fully reconstructed, setting upper limits on this anomalous branching ratio $BR(t \to Zq) < 7 \times 10^{-4}$ which are the most stringent as of today\,\cite{CMS PAS TOP-12-037}. The $tZq$ coupling can be tested also on single top quark plus $Z$ boson events, where they would appear as an excess of trlepton plus jets events but with different kinematics with respect to the previously discussed analysis. This analysis is performed by the CMS collaboration with the full 2011 dataset corresponding to 4.9\,fb$^{-1}$ using 7\,TeV of data. The corresponding upper limits are $BR(t \to Zq) <  51 \times 10^{-4}$ using single top events\,\cite{CMS-PAS-TOP-12-021}; the results could be greatly improved using the full dataset, and combined with the ones extracted from $t \bar t$ events, in order to further increase sensitivity. A possible anomalous coupling of special interest is the one of the top quark with the Higgs boson, through $tHc$ coupling. ATLAS analyzes top-antitop quark production where one of the two top quarks decays into an Higgs boson and a quark\,\cite{ATLAS-CONF-2013-081}. By focuing on Higgs decays to photons, the top quark that decays through FCNC can be fully reconstructed with minimal ambiguities. In absence of an excess in the diphoton invariant mass peak, 95\% CL upper limits are set on the corresponding branching ratio $BR(t \to Hq) < 17  \times 10^{-4}$. All the above constraints on branching ratios can be translated to constraints on the such anomalous coupling, as shown in the respective papers.

\section{The top quark and the newly discovered boson}

The discovery in 2012 of a new heavy boson\,\cite{Chatrchyan:2012ufa,Aad:2012tfa} at the LHC represents an historical milestone in our understanding of nature. To examine whether this particle truly plays the role of the electroweak symmetry breaking (ESB) mechanism agent, it is crucial to study its coupling to all known particles. The Higgs boson has been discovered mainly through its direct coupling with the other known heavy bosons ($W/Z$) and only indirectly with fermions through loops. A multitude of new physics scenarios could be hiding in those particle loops. The Yukawa structure of the coupling of the Higgs to fermions is largely unexplored: as of today, we only have mild suggestions of the Higgs coupling to $b$ quarks\,\cite{Aaltonen:2012qt} and to $\tau$ leptons\,\cite{CMStau}. Studying the direct coupling of Higgs to fermions through the associated production of top quarks together with the new heavy boson could lead to the direct measurement of the top Yukawa coupling, and thus to possible deviations from the SM predictions in the top-Higgs interaction as foreseen by natural new physics scenarios. In fact, the SM theory appears unnaturaly fine-tuned. This unnatural fine-tuning would be removed by the existence of exotic partners of the top quark that could be either of fermionic nature, such as those predicted by Composite Higgs and Little Higgs, or of bosonic nature, such as the supersymmetric top partner. In particular, both Composite/little Higgs and SUSY would predict final states that would largely overlap with the ones needed for studying top and Higgs production in the SM, i.e. one or more top quarks, in addition to one or more Higgs bosons.

The search for $t \bar t H$ production is experimentally very challenging for a multitude of reasons. First, the theoretical NNLO prediction for this process at 8\,TeV collisions amounts to only 130fb\,\cite{Dawson:2003zu}, i.e. approximately one for every $4 \cdot 10^{11}$ LHC collisions; only about one every 200 Higgs events is produced in association with top quarks. Also, both top quarks and Higgs bosons are extremely short-lived. These searches assume that the Higgs boson decays (and its decay rates) are the ones predicted by the SM. It is thus expected to detect two-to-four particles coming from the Higgs decay, and six particles coming from the top-antitop quark system decay, leading to some of the busiest events under study in high energy physics. In order to reach sensitivity to such a small and complex signal, the search for $t \bar t H$ production mandates a careful choice of only a handful of the allowed signatures. Backgrounds are generally very large as the signature is dominated by the presence of the top-antitop pair, and the cross section for $t \bar t+X$ production is three orders of magnitude larger.

Both ATLAS and CMS analyzed $t \bar t H$ events where $H \to b \bar b$; the analyses proceed similarly: first, the sample is split into events with exactly one high $p_T$ lepton - the semileptonic top-antitop decays - or two high $p_T$ leptons - the dileptonic top final state. Both samples are further split into multiple subsamples characterized by different jet multiplicities. The Higgs boson mass is reconstructed within combinatorial ambiguities whenever a sufficient number of objects are identified. The signal is discriminated from the dominant $t \bar t$+jets background by mean of BDTs that exploit the peculiar kinematical and topological characteristics of the top-antitop-Higgs production\,\cite{tthbbtt,ATLAS:2012cpa}. CMS also studied events where the Higgs decays to taus; here BDTs are used again, where the discrimination power comes from combining observables associated to the quality of the reconstructed taus. For all of the above channels, the sensitivity to the signal is estimated by a likelihood fit over the binned BDT distributions\,\cite{tthbbtt}.

The excellent resolution of the CMS and ATLAS electromagnetic calorimeters allows the reconstruction of a very narrow Higgs peak. To increase sensitivity and maximize acceptance, all $t \bar t$ system decays are collected, split into events with at least five high $p_T$ jets (hadronic channel), and events with at least two high-$p_T$ jets and one high-$p_T$ lepton (leptonic channel). In both instances at least one jet is required to be $b$-tagged. All jets are required to be sufficiently energetic as to suppress the contamination of the dominant Higgs production process. The contribution of single top plus Higgs production\,\cite{Biswas:2013xva} has been estimated to be small. The sensitivity to the signal is estimated by a likelihood fit over the diphoton invariant mass distribution\,\,\cite{tthgg,ATLAStthgg}.

\begin{figure*}
  \begin{center}
    \includegraphics[width=6cm]{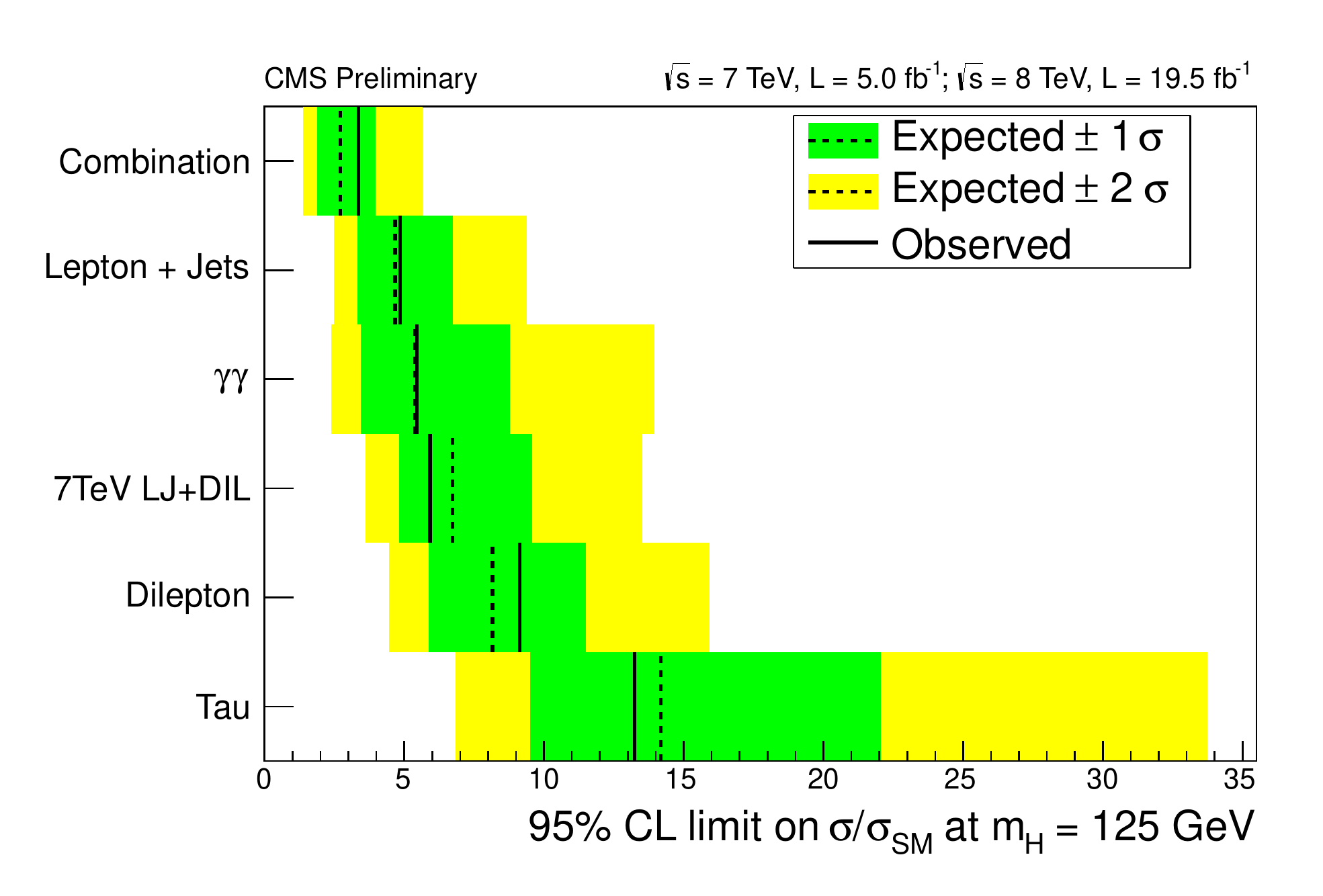}
       \includegraphics[width=6cm]{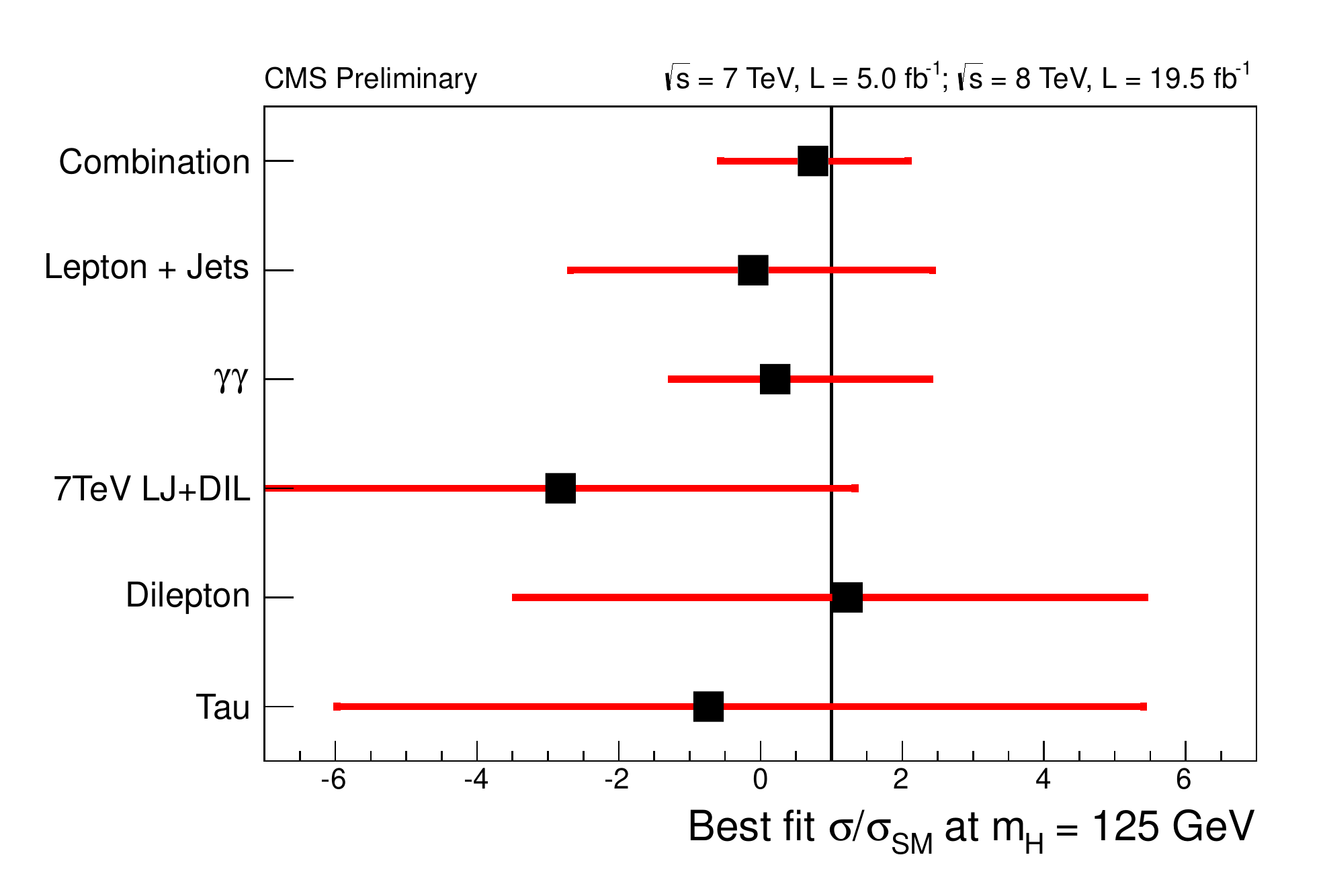}
    \caption{\label{fig:tth} The observed and expected 95\% CL upper limits on $t \bar t H$ production (left) and best-fit values (right) of $\mu$ = $\sigma/\sigma_{SM}$ for the lepton + jets (LJ), dilepton (DIL), $\tau \tau$ and $\gamma \gamma$ channels separately from the 2012 8 TeV dataset, the combination of the lepton + jets and dilepton channels from the 2011 7 TeV dataset, and the combination of all of the channels, for $m_{H} = 125$\,GeV.}
  \end{center}
\end{figure*}

The most precise current result is obtained by the combination of the several CMS results. The combination of all the above channels is accomplished using the same techniques employed in the global CMS Higgs combination\,\cite{Chatrchyan:2012ufa}. Figure\,\ref{fig:combo}, left side, shows the expected and observed limit from the individual analyses and from their combination, for an assumed Higgs boson mass of 125\,GeV. Combining these analyses improves the expected limit by 34\% compared to the best individual result for a Higgs mass of 125\,GeV. Figure\,\ref{fig:tth}, right side, shows the best fit signal strength $\mu$ for each channel contributing to the combination. For the combination of all $t \bar t H$ channels, the median expected limit for $m_{H}$ = 125\,GeV is $2.7 \times \sigma_{SM}$ while the observed limit is $3.4 \times \sigma_{SM}$, and the best-fit value for $\mu$ is $0.74\pm1.34$ (68\% CL).

The sensitivity of LHC experiments to $t \bar t H$ production has quickly increased over the past few months; it is thus possible that the first direct determination of the Yukawa coupling lies just around the corner.

\section{Conclusions}

An overview of recent results on measurements of top quark properties performed by the CDF, D0, ATLAS and CMS collaborations has been shown at this conference. The Tevatron experiments almost concluded providing their final results using the full Tevatron dataset, while LHC collaborations showed a number of impressive results that take only partly advantage of the much larger integrated dataset. It will thus be exciting to hear about the new top quark results in the next  $\sim 18$ months that separate us from the reopening of the LHC operations. A particular emphasis has been placed on the special relation between the top quark and the Higgs boson, both in the context of the Standard Model and in complete or effective theories predicting additional Higgs bosons or additional interactions between the neutral Higgs boson and the top quark. The LHC will bring a top quark dataset approximately $1000 \times$ larger during the next years, that will allow far deeper understanding of the top quark properties.

\section*{Acknowledgments}

The author wishes to thank the CDF, D0, ATLAS and CMS collaborations for providing the excellent results discussed in this paper, and Reinhard Schwienhorst for the useful discussions. The authors would like also to thank the conference organizers for the excellent organization and the pleasant atmosphere.


\begin{thebibliography}{0}    


\bibitem{Abe:1995hr}
  F.~Abe {\it et al.}  [CDF Collaboration],
  Phys.\ Rev.\ Lett.\  {\bf 74} (1995) 2626
  
\bibitem{Abachi:1995iq} 
  S.~Abachi {\it et al.}  [D0 Collaboration],
  Phys.\ Rev.\ Lett.\  {\bf 74}, 2632 (1995)
  
\bibitem{Chatrchyan:2012ufa} 
  S.~Chatrchyan {\it et al.}  [CMS Collaboration],
  Phys.\ Lett.\ B {\bf 716}, 30 (2012)
  arXiv:1207.7235
  
\bibitem{Aad:2012tfa} 
  G.~Aad {\it et al.}  [ATLAS Collaboration],
  Phys.\ Lett.\ B {\bf 716}, 1 (2012)
  arXiv:1207.7214

  
\bibitem{Baernreuther:2012ws} 
  P.~Baernreuther, M.~Czakon and A.~Mitov,
  Phys.\ Rev.\ Lett.\  {\bf 109}, 132001 (2012)
  arXiv:1204.5201


\bibitem{Reinhard}
R.~Schwienhorst, these proceedings



\bibitem{Baak:2012kk} 
  M.~Baak, M.~Goebel, J.~Haller, A.~Hoecker, D.~Kennedy, R.~Kogler, K.~Moenig and M.~Schott {\it et al.},
  Eur.\ Phys.\ J.\ C {\bf 72}, 2205 (2012)
  arXiv:1209.2716
  
\bibitem{Degrassi:2012ry} 
  G.~Degrassi, S.~Di Vita, J.~Elias-Miro, J.~R.~Espinosa, G.~F.~Giudice, G.~Isidori and A.~Strumia,
  JHEP {\bf 1208}, 098 (2012)
  arXiv:1205.6497

\bibitem{ATLAS-CONF-2013-046}
G.~Aad {\it et al.}  [ATLAS Collaboration],  ATLAS-CONF-2013-046

\bibitem{TevCombo} 
  Tevatron Electroweak Working Group, CDF and D0 Collaborations,
  arXiv:1305.3929
  
\bibitem{Chatrchyan:2013xza} 
  S.~Chatrchyan {\it et al.}  [CMS Collaboration],
  arXiv:1307.4617

\bibitem{Aaltonen:2013kna} 
  T.~A.~Aaltonen {\it et al.}  [CDF Collaboration],
  Phys.\ Rev.\ Lett.\ 
  [Phys.\ Rev.\ Lett.\  {\bf 111}, 202001 (2013)]
  [arXiv:1308.4050 [hep-ex]].
 
\bibitem{CMS-PAS-TOP-12-030}
  S.~Chatrchyan {\it et al.}  [CMS Collaboration], CMS-PAS-TOP-12-030

\bibitem{CMSpole} 
  S.~Chatrchyan {\it et al.}  [CMS Collaboration],
  arXiv:1307.1907
  
  \bibitem{CMS PAS-12-031}
  S.~Chatrchyan {\it et al.}  [CMS Collaboration], CMS PAS-12-031
  
\bibitem{CMS-PAS-B2G-12-023}
 S.~Chatrchyan {\it et al.}  [CMS Collaboration], CMS-PAS-B2G-12-023
  

\bibitem{Aaltonen:2011kc} 
  T.~Aaltonen {\it et al.}  [CDF Collaboration],
  Phys.\ Rev.\ D {\bf 83}, 112003 (2011)
  arXiv:1101.0034

\bibitem{Aaltonen:2013vaf} 
  T.~A.~Aaltonen {\it et al.}  [CDF Collaboration],
  Phys.\ Rev.\ D {\bf 88}, 072003 (2013)
  arXiv:1308.1120


\bibitem{D0 CONF-6394}
V.~M.~Abazov {\it et al.}  [D0 Collaboration], D0 CONF-6394

\bibitem{Abazov:2013wxa} 
  V.~M.~Abazov {\it et al.}  [D0 Collaboration],
  arXiv:1308.6690

\bibitem{CMS-PAS-TOP-12-004}
S.~Chatrchyan {\it et al.}  [CMS Collaboration],  CMS-PAS-TOP-12-004

\bibitem{Chatrchyan:2012cxa} 
  S.~Chatrchyan {\it et al.}  [CMS Collaboration],
  Phys.\ Lett.\ B {\bf 717}, 129 (2012)
  [arXiv:1207.0065 [hep-ex]].

\bibitem{ATLAS-CONF-2013-078}
G.~Aad {\it et al.}  [ATLAS Collaboration],  ATLAS-CONF-2013-078

\bibitem{ATLAS-CONF-2012-057}
G.~Aad {\it et al.}  [ATLAS Collaboration],  ATLAS-CONF-2012-057



\bibitem{Aad:2013hla} 
  G.~Aad {\it et al.}  [ATLAS Collaboration],
  Eur.\ Phys.\ J.\ C {\bf 73}, 2465 (2013)
  arXiv:1302.3694
  
\bibitem{CMS-PAS-HIG-12-052}
S.~Chatrchyan {\it et al.}  [CMS Collaboration],  CMS-PAS-HIG-12-052
  
\bibitem{Aad:2012tj} 
  G.~Aad {\it et al.}  [ATLAS Collaboration],
  JHEP {\bf 1206}, 039 (2012)
  arXiv:1204.2760
  
\bibitem{Abazov:2008rn} 
  V.~M.~Abazov {\it et al.}  [D0 Collaboration],
  Phys.\ Rev.\ Lett.\  {\bf 102}, 191802 (2009)
  [arXiv:0807.0859 [hep-ex]].


\bibitem{ATLAS-CONF-2013-063}
G.~Aad {\it et al.}  [ATLAS Collaboration], ATLAS-CONF-2013-063

\bibitem{CMS PAS TOP-12-037}
S.~Chatrchyan {\it et al.}  [CMS Collaboration], CMS PAS TOP-12-037

\bibitem{CMS-PAS-TOP-12-021}
S.~Chatrchyan {\it et al.}  [CMS Collaboration], CMS-PAS-TOP-12-021

\bibitem{ATLAS-CONF-2013-081} 
G.~Aad {\it et al.}  [ATLAS Collaboration],  ATLAS-CONF-2013-081




\bibitem{Aaltonen:2012qt} 
  T.~Aaltonen {\it et al.}  [CDF and D0 Collaborations],
  Phys.\ Rev.\ Lett.\  {\bf 109}, 071804 (2012)
  arXiv:1207.6436
  
\bibitem{CMStau} 
S.~Chatrchyan {\it et al.}   [CMS Collaboration],
CMS-HIG-13-004


\bibitem{Dawson:2003zu} 
  S.~Dawson, C.~Jackson, L.~H.~Orr, L.~Reina and D.~Wackeroth,
  Phys.\ Rev.\ D {\bf 68}, 034022 (2003)
  hep-ph/0305087


    
  
  
\bibitem{tthbbtt} S.~Chatrchyan {\it et al.}  [CMS Collaboration], 
CMS-HIG-13-019

\bibitem{ATLAS:2012cpa} 
G.~Aad {\it et al.}   [ATLAS Collaboration],
  ATLAS-CONF-2012-135.

\bibitem{Biswas:2013xva} 
  S.~Biswas, E.~Gabrielli, F.~Margaroli and B.~Mele,
  JHEP {\bf 07}, 073 (2013)
  arXiv:1304.1822

\bibitem{tthgg} S.~Chatrchyan {\it et al.}  [CMS Collaboration], 
CMS-HIG-13-015 

\bibitem{ATLAStthgg}
G.~Aad {\it et al.}   [ATLAS Collaboration], ATLAS-CONF-2013-080


\end{thebibliography}
\end{document}